\documentclass[twocolumn,twoside,slac_two]{revtex4}
\usepackage{graphicx}
\usepackage{fancyhdr}
\newtoks\nslashfraction
 \nslashfraction={.05}
 \newcommand{\nslash}[1]{\setbox0\hbox{$ #1 $}
   \setbox0\hbox to \the\nslashfraction\wd0{\hss \box0}/\box0 }

\begin{document}

\title{Discovery Potential for Di-lepton and Lepton+Etmiss Resonances at High Mass with ATLAS}

%

\author{M.I. Pedraza-Morales}
\affiliation{Department of Physics, University of Wisconsin, Madison, WI 53706, USA}
\begin{abstract}
This paper describes the discovery potential for new resonances with the ATLAS experiment. The resonances discussed in here are the $Z'$, leptoquarks, graviton and $W'$ resonances in some of their leptonic final states, considering a center-of-mass energy of $\sqrt{s}=14$~TeV for all of them,  and an estimation of the potential of the $W'$ search at the early center-of-mass energy of the LHC . The studied scenarios show that an initial run of few tens of pb$^{-1}$ would be enough to go beyond the current limits in most of these models.  
\end{abstract}

\maketitle

\thispagestyle{fancy}


\section{Introduction}
Final states containing two leptons, or one lepton and missing transverse energy offer a great potential for early discoveries at the LHC. They can be generated by several kinds of resonances. All the results presented here are done with Monte Carlo samples simulated using a complete description of the ATLAS detector, and are based on~\cite{Aad:2009wy}. 

ATLAS detector subsystems relevant for these searches are the tracking system (inner detector), the electromagnetic calorimeter, the  hadronic calorimeter, and the muon spectrometer. The physical objects used in this analysis are built as follows: Muons are reconstructed independently in the inner detector and the muon spectrometer, and are combined using statistical combination; electrons are reconstructed from clusters of calorimeter cells with a matched track in the inner detector; and the missing transverse energy, $\mbox{ }\nslash E_T$ , is calculated from the energy collected in the calorimeter cells.  

\section{Lepton and Missing Transverse Energy Final States}

\subsection{$W'$}
 
\begin{figure}[h]
\centering
\includegraphics[width=40mm]{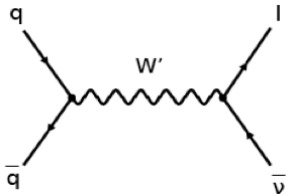}
\caption{Feynman diagram for $W'$ production.} \label{Wp_Feynman}
\end{figure}

New heavy gauge bosons, like $W'$, are predicted in various extensions of the Standard Model (SM). The production cross section depends on the $W'$ mass, and in some models is large enough for $m_{W'} \leq \sqrt{s}/2$ that a $W'$ boson becomes a candidate for early discoveries at the LHC~\cite{PDG}. 
The D$\O$ experiment has set the present lower limit for the $W'$ boson mass to $m_{W'} > 1$~TeV at $95\%$ C.L.~\cite{D0:2007bs}. 
Studies presented here are based on Altarelli Reference Model~\cite{Altarelli:1989ff}, also known as Sequential Standard Model (SSM): in this model, the new heavy gauge boson $W'$ couples to fermions in the same way as the $W$ boson in the SM. 

The leptonic final state of a $W'$, Fig.~\ref{Wp_Feynman},  is given by a lepton and a neutrino. The momentum information of the neutrino can be inferred only partially from the energy imbalance in the detector, which gives the missing transverse energy, $\nslash E_T$, of the event. In this section the analyses for muon and electron channels are discussed.

\subsubsection{Event Selection}

\begin{figure}[h]
\centering
\includegraphics[width=75mm]{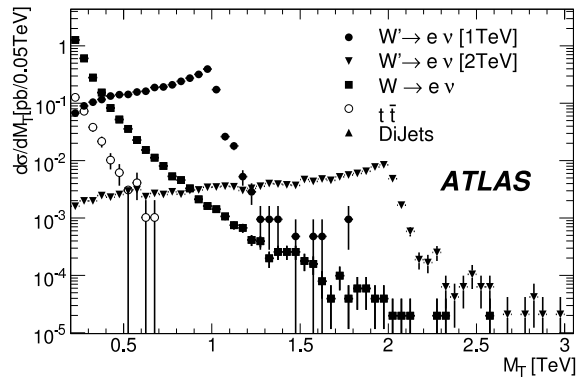}
\caption{Transverse mass spectrum of the $W' \rightarrow e \nu$ signals and SM backgrounds normalized to the cross section after selecrion for $1$~pb$^{-1}$ of integrated luminosity.} \label{Wp_eleTmass}
\end{figure}

The decay $W' \rightarrow \ell \nu$ provides a clean signature in the transverse mass spectrum. The largest backgrounds are the high$-p_{T}$ tail of the $W \rightarrow \ell \nu$ decays and $t \bar{t}$ production. In addition the background coming from fake leptons, multi-jet production, is also considered. 

The following basic cuts were applied: just one isolated electron or muon with $p_{T}> 50$~GeV and $|\eta|<2.5$; and  $\nslash E_{T}>50$~GeV. In order to separate the signal from the reducible background additional cuts are applied:  the scalar sum of the $p_{T}$ of tracks in a cone of $dR=0.3$ around the lepton divided by lepton $p_T$ must be less than $0.05$ , and the ratio $\sum p_{T}^{leptons}/( \sum p_{T}^{leptons} + \sum E_{T})$, that gives the fraction of energy attributed to the leptons (including neutrinos, which are assumed to be the main contribution to $\nslash E$), more than $0.5$. These two last cuts are required in order to reject events with lots of additional activity from jets (like $t\bar{t}$ and di-jets).  

Reconstructing the transverse mass of the event as:
\begin{equation}
m_T= \sqrt{2p_T^{lepton} \nslash E_T (1 - cos \Delta \phi_{lepton,\mbox{  } \nslash E_T})},
\label{transvermass}
\end{equation}
\noindent Fig.~\ref{Wp_eleTmass} shows the transverse mass spectrum after all cuts for the signal, $W' \rightarrow e \nu$ with $m =1, 2$~TeV,  and for the background.  

\subsubsection{Discovery Potential}
\begin{figure}[h]
\centering
\includegraphics[width=75mm]{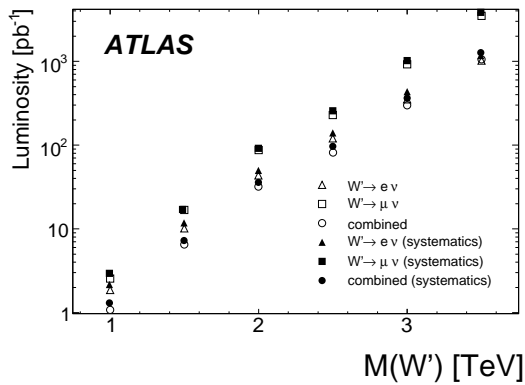}
\caption{Integrated luminosity needed for a $5\sigma$ excess for $W'\rightarrow e \nu, \mu \nu$ .} \label{Wp_Lumi}
\end{figure}

In order to estimate the ATLAS discovery potential in the search for a $W'\rightarrow \ell + \nslash E_T$ signal, the luminosity needed for a $5\sigma$ excess is obtained as a function of the mass of the $W'$ boson.
The significance is obtained from the expected number of signal and background events in the region $m_{T}>0.7m_{W'}$. Calling these expected numbers $s$ and $b$, respectively, the significance $S$ is obtained as
\begin{equation}
S= \sqrt{2((s+b)ln(1+s/b)-s)},
\label{significance}
\end{equation}
\noindent which gives a good approximation to the likelihood-ratio based significance in the low statistics regime. Fig.~\ref{Wp_Lumi} shows the expected integrated luminosity needed for a $5\sigma$ excess as a function of the mass of the $W'$ boson.
This study shows that, even with a total integrated luminosity as low as $10$~pb$^{-1}$, it would be possible to go beyond the limits set by previous searches of this type of boson.

\section{Di-lepton Final States}
New heavy states forming narrow resonances decaying into di-leptons are predicted in many extensions of the SM. In this section, the analysis for  three such resonances are described. Due to the simplicity of the final state, these channels will be very important for the early physics results of ATLAS.    

\subsection{$Z'$}
\begin{figure}[h]
\centering
\includegraphics[width=40mm]{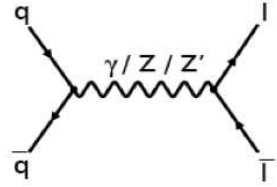}
\caption{Feynman diagram for the $Z'$ production.} \label{Zp_Feynman}
\end{figure}
Several models predict the existence of additional neutral gauge bosons. Since the experimental consequences are very similar in the di-lepton final state, just a few representative models are studied: the SSM, the $E_6$ and the Left-Right Symmetric models~\cite{Cvetic:1991kh}. In Table~\ref{zp_exclusion} the $95\%$ C.L. exclusions are shown for different $Z'$. The leptonic final state of a $Z'$, as seen in Fig.~\ref{Zp_Feynman},  is given by two opposite charge high $p_{T}$ leptons.
\begin{table}[h]
\begin{center}
\caption{Status of $95\%$ C.L. limits on various $Z'$ models ~\cite{Alcaraz:2006mx} ~\cite{PDG} ~\cite{Aaltonen:2008ah}.}
\begin{tabular}{|l|c|c|c|}
\hline 
\textbf{$Z'$ Model} & \multicolumn{2}{c|}{ \textbf{Indirect}} & \textbf{Direct } \\
                  & \multicolumn{2}{c|}{\textbf{Searches (GeV)}} & \textbf{Searches (GeV)} \\
\cline{2-3}
                  & \textbf{Electroweak} & \textbf{LEP}  & \textbf{$p \bar{p}$ Colliders} \\
\hline $Z'_{\chi}$  & $680$ &   $781$ &   $892$ \\
\hline $Z'_{\psi}$  & $137$ &   $481$ &   $878$ \\
\hline $Z'_{\eta}$  & $619$ &   $515$ &   $904$ \\
\hline $Z'_{LRSM}$  & $860$ &   $804$ &   $-$  \\
\hline $Z'_{SSM}$   & $1500$ & $1787$ &  $1030$  \\
\hline
\end{tabular}
\label{zp_exclusion}
\end{center}
\end{table}

\subsubsection{Event Selection}
\label{esel_zp}

Section ~\ref{esel_zpA} presents muon and electron cases first, due to the similarity in their selection criteria, then, ~\ref{esel_zpB} describes the $\tau$ channel, which requires more discussion. 

\paragraph{$Z' \rightarrow ee, \mu \mu$}
\label{esel_zpA}

The decay $Z' \rightarrow ee, \mu \mu$ provides a clean signature in the mass spectrum. The largest background is the irreducible Drell-Yan process. The contribution of the other studied backgrounds is less than $30\%$ of the Drell-Yan cross-section. The study presented therefore considers only the Drell-Yan background. 

\begin{figure}[h]
\centering
\includegraphics[width=80mm]{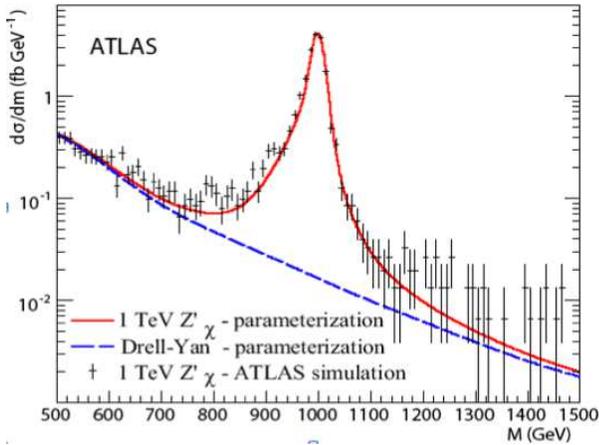}
\caption{Mass spectrum for a $m=1$~TeV $Z'_{\chi} \rightarrow e^{+}e^{-}$ obtained with ATLAS full simulation (histogram) and the parameterization (solid line). The dashed line corresponds to the parameterizations of the Drell-Yan process.} \label{Zp_Mass}
\end{figure}

The following cuts were applied: $|\eta|<2.5$, two electrons(muons) in the event, at least one electron(muon)  with $p_{T}> 65$~GeV ($p_{T}> 30$~GeV) and opposite charge. Fig.~\ref{Zp_Mass} shows the mass spectrum after all cuts for a $Z'$ with $m =1$~TeV,  and the parameterization for the Drell-Yan process.

\paragraph{$Z' \rightarrow \tau \tau$}
\label{esel_zpB}

The di-tau signature is an important component to the high mass resonance search. In particular, there are models in which a hypothetical new resonance couples preferentially to the third generation. The di-tau final state can be divided into three final states: hadron-hadron (where both taus decay hadronically), hadron-lepton (where one $\tau$ decays leptonically and one decays hadronically), and lepton-lepton (where both taus decay leptonically). Here the hadron-lepton final state is considered.

\begin{figure}[h]
\centering
\includegraphics[width=70mm]{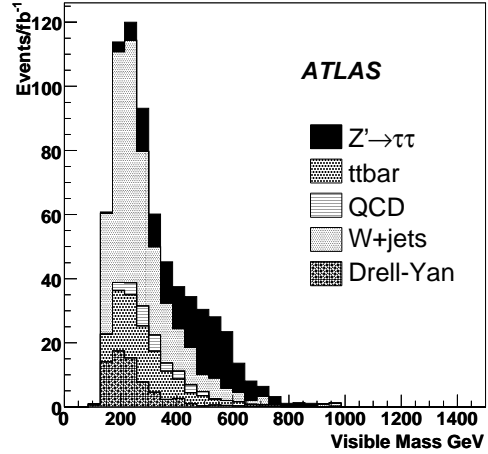}
\caption{The visible mass distribution in the $Z' \rightarrow \tau \tau \rightarrow \ell h$ analysis for signal and background processes($1$ $fb^{-1}$ of data is assumed).} \label{Zp_Mass_tau}
\end{figure}

For this channel, the background is large and includes Drell-Yan, W+jets, $t \bar{t}$ and  multi-jet production.

The basic cuts for this channel are: $\tau$ candidate with $p_{T}>60$~GeV, an electron(muon) with $p_{T}> 27$~GeV ($p_{T}>22$~GeV), and $\nslash E_{T} > 30$ . After these initial cuts, several additional requirements are needed in order to maximize the expected signal significance: the scalar sum of the $p_{T}$ of tracks in a cone of $0.2$ around the lepton divided by lepton $p_T$ must be less than $0.1$, the leptons must have opposite charge, the transverse mass using the lepton kinematics and the  $\nslash E_{T}$ less than $35$~GeV, the total $p_{T}$ less than $70$~GeV, and the $m_{vis}$ must be greater than $300$~GeV.

In the case of the lepton-hadron channel it is not possible to get directly the invariant mass since there is lost energy due to the neutrinos. Thus it is helpful to define the visible mass:


\vspace{-4mm}
\begin{equation}
m_{vis}= \sqrt{(p_{\ell}+p_{h}+\nslash p_T)^2}
\label{visible}
\end{equation}

Where $\nslash p_T$ is the four-vector for the missing transverse energy, defined as $\nslash p_T = (\nslash E_{T_X}, \nslash E_{T_y}, 0, |\nslash E_{T}|)$. In Fig.~\ref{Zp_Mass_tau} the visible mass spectrum for signal and background is shown, for a $Z'_{SSM}$ with $m_{Z'}=600$~GeV. 





\subsubsection{Discovery Potential}

In this section the different expectations for the discovery potential  of a $Z'$ in the leptonic channel are presented.

\paragraph{$Z' \rightarrow ee, \mu \mu $}

\begin{figure}[h]
\centering
\includegraphics[width=80mm]{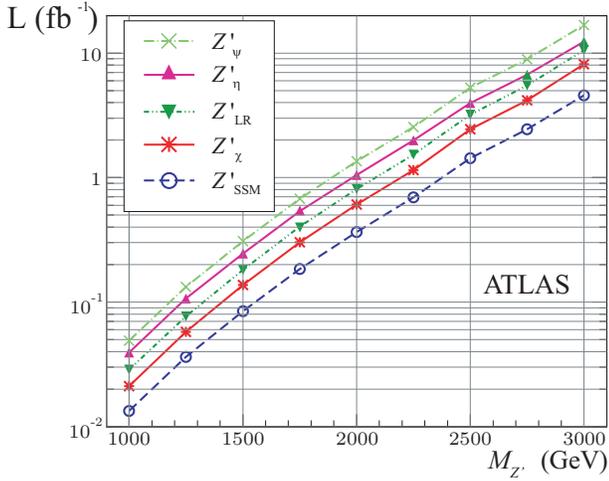}
\caption{Integrated luminosity needed for a $5\sigma$ discovery of $Z' \rightarrow ee $ as a function of the $Z'$ mass.} \label{Zpele_Lumi}
\end{figure}

To calculate the significance in the case of the $Z' \rightarrow ee $, a parameterization  of the mass spectrum of the signal and background was used. Fig.~\ref{Zpele_Lumi} shows the integrated luminosity needed for a $5 \sigma$ discovery of the usual benchmark $Z'$ models as a function of the $Z'$ mass for di-electron channel. Only statistical uncertainties were taken into account. 

From the plot we can see that less than $100$~pb$^{-1}$ are needed  to discover a $1$~TeV $Z'$, and $1fb^{-1}$ for a $2$~TeV $Z'$.

\begin{figure}[h]
\centering
\includegraphics[width=80mm]{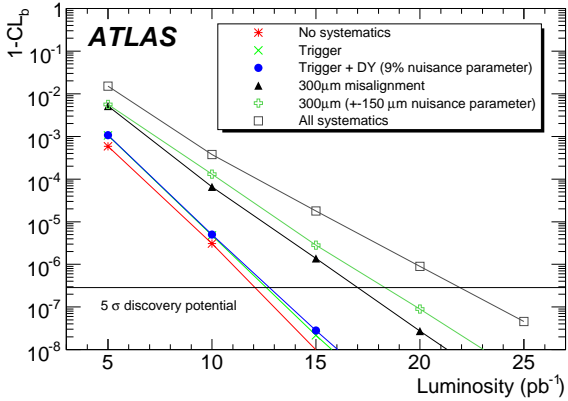}
\caption{Results of $1-CL_b$ for $m=1$~TeV $Z'_{SSM} \rightarrow \mu \mu$ boson. The horizontal line indicates the $1-CL_b$ value corresponding to $5\sigma$} \label{Zpmuo_Lumi}
\end{figure}

Fig.~\ref{Zpmuo_Lumi} shows $1-CL_b$, where $1-CL_b$ can be seen as the probability that the background gives an output that looks like the signal,  as a function of the integrated luminosity for a $Z'_{SSM}$ boson with $m=1$~TeV, for di-muon channel. For high$-p_{T}$ muons ($>300$~GeV) the muon spectrometer alignment is the main contribution to the signal sensitivity degradation. Between $0$ and $300$ $\mu$m misalignment, the luminosity needed to reach $5\sigma$ discovery increase of $42\%$ (see Fig.~\ref{Zpmuo_Lumi}).

From  Fig. ~\ref{Zpmuo_Lumi}, we can see that with less than $100$~pb$^{-1}$ of data a $Z'$ in the muon channel could give an excess of $5\sigma$.

\paragraph{$Z' \rightarrow \tau \tau $}

\begin{figure}[h]
\centering
\includegraphics[width=80mm]{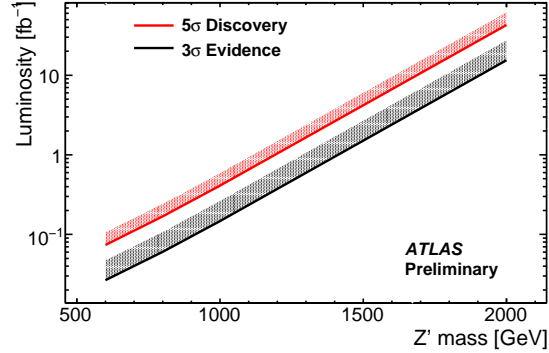}
\caption{Integrated luminosity required for $3\sigma$ evidence and a $5\sigma$ discovery as a function of the true $Z'_{SSM}$ mass, for $Z'_{SSM} \rightarrow \tau \tau$ ~\cite{Czyczula}.} \label{Zptau_Lumi}
\end{figure}

The potential of ATLAS to discover a high-mass resonance decaying into two $\tau$ leptons, as defined by Eq.~\ref{significance}, is presented in the Fig.~\ref{Zptau_Lumi}. It shows the amount of luminosity required for $3\sigma$ evidence or for a $5\sigma$ discovery as a function of the true mass of the $Z'$. 

It should be noted that for models where $Z'$ couples preferentially to the third generation  the limits are lower than the ones presented in the Table~\ref{zp_exclusion}, therefore it was considered important to look at the lower invariant mass region in this channel.

If a $Z'$ with a relatively low mass is coupled to the third family, it could be discovered in the di-tau channel with a few hundred pb$^{-1}$ of data.

\subsection{Technicolor}

\begin{figure}[h]
\centering
\includegraphics[width=50mm]{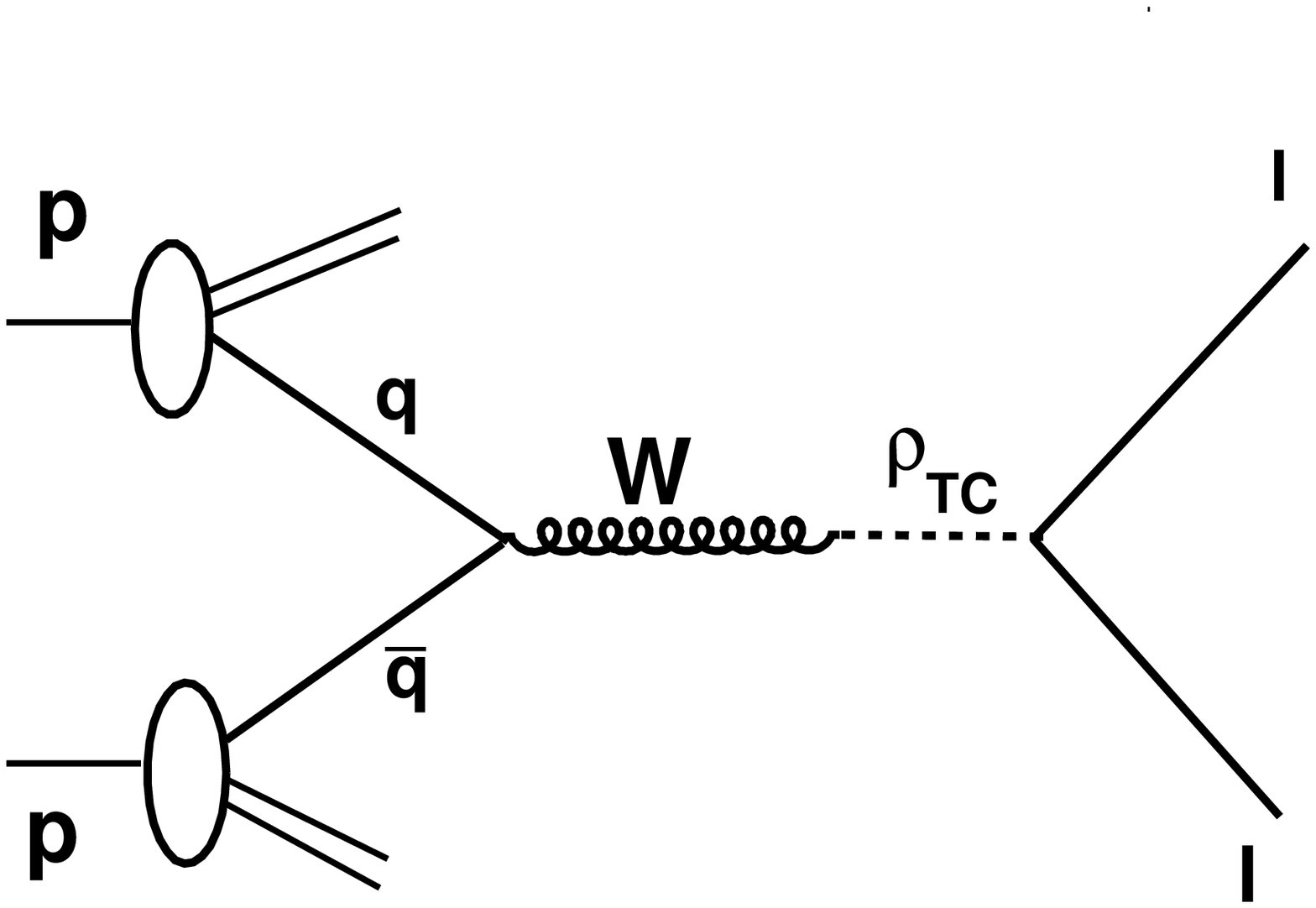}
\caption{Feynman diagram for technimeson production.} \label{Tch_Feynman}
\end{figure}

Some technicolor models predict new technihadron states that can be produced at the LHC. One of the most promising channels to search for these new hadrons is the di-lepton decay of the $\rho_{TC}$ and $\omega_{TC}$, as depicted in Fig.~\ref{Tch_Feynman}. The ``Technicolor Strawman Model'' or TCSM is used as a benchmark model. 
The limits set by CDF rule out  $\rho_{TC}$ and $\omega_{TC}$ masses below $280$~GeV for a particular choice of the TCSM parameters. This section presents the muon channel. The main background for this search is the Drell-Yan process.

\subsubsection{Event Selection}

\begin{figure}[h]
\centering
\includegraphics[width=80mm]{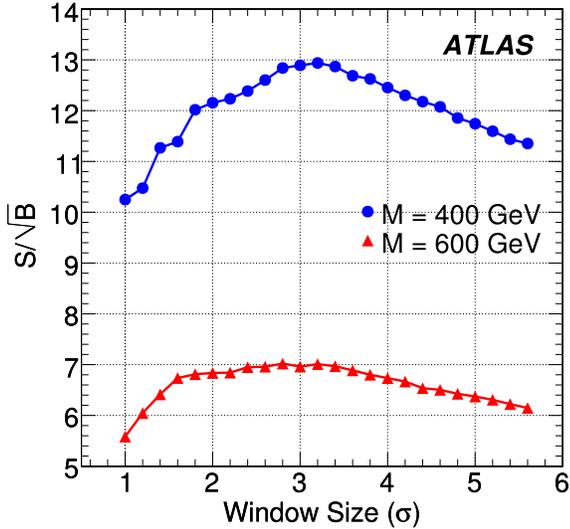}
\caption{For two different $\rho_{TC}$, $\omega_{TC}$ signal masses, $s/ \sqrt{b}$ is plotted as a function of mass-window size for windows centered on the peak mass for decays to $\mu \mu$.} \label{technicolor_mw}
\end{figure}

The basic cuts for $\rho_{TC}$ and $\omega_{TC}$ in the muon channel are: $|\eta|<2.5$, two muons with  $p_{T}>30$~GeV, and opposite charge. Since the best search sensitivity is not obtained by examining the entire dimuon mass distribution, a mass window around a given mass peak is optimized based on the signal significance. 

Fig.~\ref{technicolor_mw} shows the value of the significance for two different mass points as a function of the mass window. From the plot is evident that it is optimal to use a mass window $\pm 1.5 \sigma(m)$, where $\sigma(m)$ is the width due to detector resolution since the technimeson natural widths are less than a GeV.

\subsubsection{ Discovery Potential}

\begin{figure}[h]
\centering
\includegraphics[width=80mm]{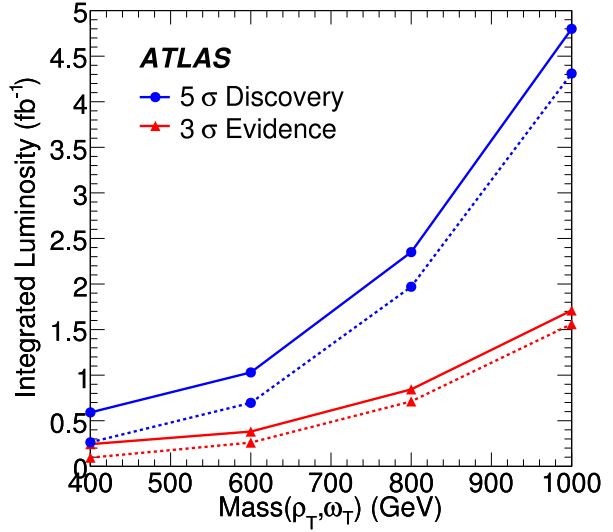}
\caption{Integrated luminosity needed for a $3\sigma$ evidence or $5\sigma$ discovery as a function of $\rho_{TC}$, $\omega_{TC}$  mass for decays to $\mu \mu$. The dashed lines include only statistical uncertainties while solid lines contain the systematic uncertainties as well.} \label{technicolor_lumi}
\end{figure}

Using number counting and the $s/\sqrt{b}$ approximation to get the significance, Fig.~\ref{technicolor_lumi} is obtained. This plot shows that  $\rho_{TC}$ or  $\omega_{TC}$ can be discovered with $600$~pb$^{-1}$, including the estimated effect of the mis-alignment of the muon detector.

\subsection{Graviton}

\begin{figure}[h]
\centering
\includegraphics[width=60mm]{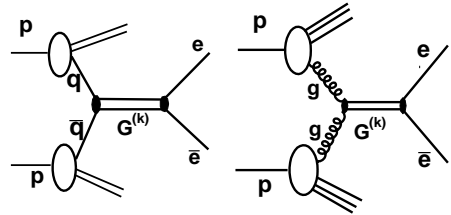}
\caption{Feynaman diagrams for a Kaluza-Klein excitation of the graviton. Additionally the graviton production has an extra con} \label{Gv_Feynman}
\end{figure}

The Randall-Sundrum model addresses the hierarchy problem by adding one extra-dimension to our four-dimensional world. It predicts the existence of a tower of Kaluza-Klein excitations of the graviton. This graviton should be observable as a resonance which decays into lepton pairs at the LHC, Fig.~\ref{Gv_Feynman}. The model parameters are the mass, $m_{G}$, and the coupling constant $\kappa/ \bar{M}_{pl}$. The current limits depend on the parameters of the model, the region of low graviton mass and larger $\kappa/ \bar{M}_{pl}$ is excluded ~\cite{Landsberg:2004mj}. Similarly to the $Z'$ and technicolor searches the main background for this resonance is the Drell-Yan process.


\subsubsection{Event Selection}

\begin{figure}[h]
\centering
\includegraphics[width=80mm]{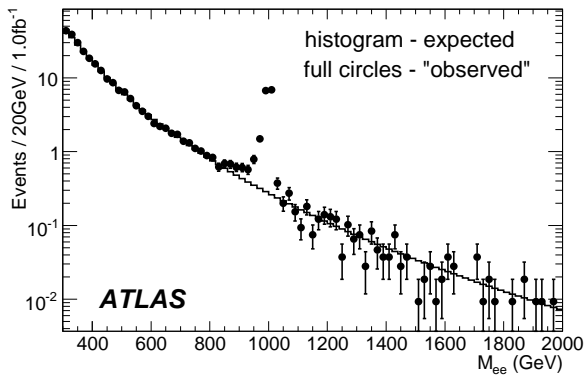}
\caption{Background  and signal+background mass spectra from full simulation for the di-electron channel.} \label{Gv_Mass}
\end{figure}

The basic cuts for this search in the electron channel are: $|\eta|<2.5$, two electrons with  $p_{T}\geq 65$~GeV, and  that the electrons are back-to-back \textit{i.e.} $cos \Delta \Phi_{ee} <0$. The parameterized Drell-Yan background and the signal are shown in Fig.~\ref{Gv_Mass} for a graviton with $m_{G}=1$~TeV and a coupling value $\kappa/\bar{M}_{pl}=0.02$.

\subsubsection{Discovery Potential}

\begin{figure}[h]
\centering
\includegraphics[width=80mm]{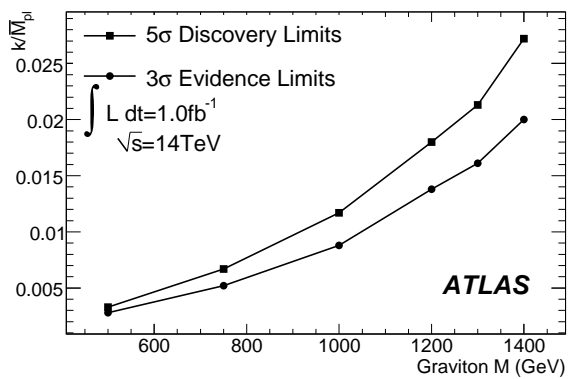}
\caption{$5\sigma$ discovery  and $3\sigma$ evidence as a function of the graviton mass in the di-electron channel.} \label{Gv_Lumi}
\end{figure}

The mass range used for graviton searches is from $300$~GeV up to $2$~TeV. The method used to get the significances is a parameterized fit approach. Fig.~\ref{Gv_Lumi} shows the $5\sigma$ discovery and $3\sigma$ evidence reach in $\kappa/\bar{M}_{pl}$ coupling constant as a function of the graviton mass for an integrated luminosity equal to $1$ $fb^{-1}$.

For some values of $k/\bar{M}_{pl}$ it is possible to have a discovery with O($100$~pb$^{-1}$).

\section{Example of the potential of early data of the LHC ($W' \rightarrow e \nu$)}

The early run of the LHC is expected to be at $\sqrt{s}=10$~TeV. With this center-of-mass energy,  new physics may already become visible. In this section, the discovery potential for $W' \rightarrow e \nu$, $\mu \nu$ is discussed. 

The lowered center-of-mass energy, at $10$~TeV, reduces the production cross section, thus the sensitivity. In the case of $W' \rightarrow e \nu$, $\mu \nu$, the signal cross section is reduced to about $50\%$, $40\%$, $35\%$, $27\%$ for $m(W') = 1, 1.5, 2, 2.5 $~TeV respectively, and about to $60\%$ for the backgrounds.

\begin{figure}[h]
\centering
\includegraphics[width=80mm]{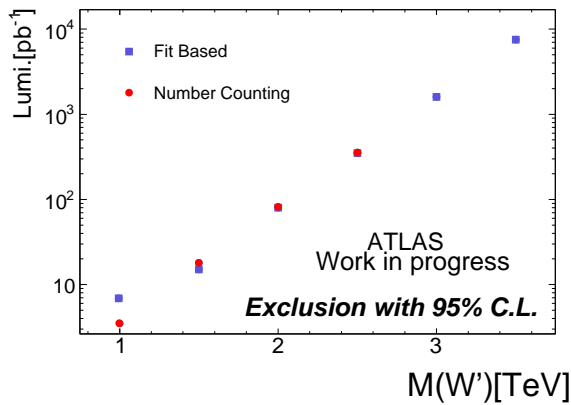}
\caption{Integrated luminosity needed for a $95\%$ C.L. exclusion for $W'\rightarrow e \nu$ as a function of the mass of the $W'$.} \label{Wp10TeV_Exclusion}
\end{figure}

The signal and the background remain without significant modifications on their shape for $\sqrt{s}=10$~TeV with respect to the ones at $\sqrt{s}=14$~TeV. Estimation for exclusion limits for this resonance was done using both number counting and a fit based-method. The preliminary estimation of the integrated luminosity needed for $95\%$ C.L. exclusion is shown in Fig. ~\ref{Wp10TeV_Exclusion}.

This results are based on a very optimistic scenario and the trigger efficiency is not applied. Nevertheless, $W'$ is one of the searches that can be performed in the first run of the LHC. With O($50$~pb$^{-1}$) of well understood data we can either discover it or exclude it beyond the current limit. Note that these results have not yet been officially approved by the ATLAS Collaboration.  

\section{Summary and Conclusions}
A selection of analysis on resonances in the di-lepton and lepton$+$Etmiss final states have been presented. LHC and ATLAS constitute a powerful tool to discover or exclude new particles. The studies based on a center of mass energy $\sqrt{s}=14$~TeV have shown that the existence of a $W'$ and $Z'$ could be established at the $5\sigma$ level even with O$(100$~pb$^{-1})$ of integrated luminosity. The initial run of few tens of pb$^{-1}$ at $10$~TeV would be enough to go beyond Tevatron limits in most of these models. 
\bigskip 

\end{document}